# Surface enhanced infrared absorption spectroscopy for graphene functionalization on copper

Irena Matulková,[1] Petr Kovaříček,[2] Miroslav Šlouf,[3] Ivan Němec,[1] Martin Kalbáč[2,*]

[1] *Faculty of Science, Charles University, Hlavova 2030/8, 128 43, Praha, Czech Republic*

[2] *J. Heyrovsky Institute of Physical Chemistry of the Academy of Sciences of the Czech Republic, Dolejškova 2155/3, 182 23 Praha, Czech Republic*

[3] *Institute of Macromolecular Chemistry, Academy of Sciences of the Czech Republic, Heyrovsky Sq. 2, 162 06 Praha, Czech Republic*

## Abstract

The monolayer form of CVD graphene is imposing crucial challenges in characterization of the targeted covalent functionalization due to the low amount of grafting moieties on the surface, which in turn hampers drawing conclusions about reactivity-properties relationships. Due to the growing interest in chemically modified graphene new, reliable and non-destructive methods for its characterization are critically required. Herein we demonstrate the use of surface-enhanced infrared absorption spectroscopy for detection of characteristic vibration modes of species being grafted to the material either via radical (Meerwein arylation) or nucleophilic substitution pathway on copper foil. The phenomenon is allowed by the appropriate metal surface morphology and no signal could have been obtained neither on Si/SiO$_2$ substrate, nor on bare copper. The surface of copper foil exhibit partial corrosion during the reaction which leads to the creation of active substrate for SEIRA. The measurements were performed using reflection-absorption and attenuated total reflection modes with almost identical results, thus making this analytical approach feasible, practical, non-destructive and easy to use for routine characterization of graphene functionalization.

[*] martin.kalbac@jh-inst.cas.cz, tel.: +420-266 05 3804

# 1  Introduction

Controlled on-demand functionalization of 2D materials is a very important milestone on the way seeking potential applications across all fields.[1–8] The strict monolayer and flat character of these materials, which is in large part responsible for their fascinating features,[9–11] also represents a critical challenge in development of appropriate characterization methods for clear, fast and unambiguous analysis of the prepared materials. The usual surface-selective techniques used in the field such as X-ray photoelectron spectroscopy[12–14] or Raman scattering[15–17] provide only limited insight into the chemical transformation taking place on the surface and typically cannot distinguish in between covalent grafting and surface contamination, which is almost inevitable due to the polymer-assisted transfer and in-solution chemical functionalization. It is thus of high importance to strive after new opportunities to obtain characteristic structural information about species being introduced to the monolayer in a non-invasive and non-destructive fashion.[18]

Recently, surface-enhanced versions of Raman spectroscopy (SERS)[19,20] and mass spectrometry (SELDI)[18] have been demonstrated to provide dramatic signal enhancement of species being covalently attached to graphene. Yet, these techniques rely on laser irradiation of small sample areas, which in some cases may lead to degradation of the sensitive samples. In addition the use of Raman spectroscopy is limited to functional groups which obey Raman selection rules.

A traditional complementary method to the Raman spectroscopy is infrared spectroscopy (IR). In addition IR on the is an absorption technique minimizing the risk of eventual damage by working with very low energy of electromagnetic radiation and have therefore been used in characterization of graphitic materials in bulk.[21–23] Therefore, the application of IR spectroscopy for characterization of functionalized graphene will be extremely beneficial. However, IR often faces low sensitivity and signal-to-background ratio problems[24] which are even more troublesome in the case of monolayer materials functionalization deposited on a substrate. This obstacle can be actually turned into an advantage if the substrate is made of a material, which is capable to induce surface-enhanced infrared absorption (SEIRA).[25] In general, the SEIRA can be induced by transition metals including Au, Ag or Cu.[26–29] Copper is particularly interesting in the context of graphene functionalization because Cu foils are routinely used for the chemical vapor deposition (CVD) synthesis of graphene. It has been shown recently that various protocols for graphene functionalization yield also roughened metal surface to induce SERS hotspots for detection of the functionalizing species.[18] Similarly, in the case of SEIRA the roughened metal surface can enhance vibrational modes of molecules with a change in dipole moment perpendicular to the surface by a factor of up to $10^4$ on Au or Ag films.[25–27] Although there are obvious benefits of the SEIRA, this method was not applied to study functional groups on graphene yet.

Herein we succeed for the first time to obtain SEIRA of functionalized graphene samples using Attenuated Total Reflection (ATR) and Reflection-Absorption IR Spectroscopy (RAIRS) techniques. The method was tested for different functional groups attached on graphene using different mechanisms, where the successful functionalization was previously confirmed by other techniques.[30] The measured spectra allowed to resolve IR absorption bands characteristic for functional groups of each particular reagent used.

## 2 Experimental part

### 2.1 Graphene synthesis and functionalization

Graphene was grown on copper using CVD, as described previously.[31,32] In brief, a polycrystalline copper foil was annealed for 20 min at 1000 °C in hydrogen atmosphere. Graphene was then grown from 1 standard cubic centimeter per minute (sccm) $CH_4$ for 45 min and then annealed for another five minutes in $H_2$ atmosphere. The sample was then cooled to room temperature. For the experiments on Si/SiO$_2$ substrate the transfer was performed using the nitrocellulose-based technique.[33]

The graphene fluorination was performed in a home-made apparatus. The sample on a substrate was placed in a vacuum chamber containing blank Si/SiO$_2$ wafer, evacuated to ≈2 × $10^{-4}$ mbar, then isolated from the pump and connected to the solid XeF$_2$ (Aldrich, 99.99%) reservoir. When the pressure reached 8 mbar (about 60 s), the valve to the reservoir was closed and the system was evacuated to remove residual XeF$_2$ vapors and then slowly ventilated to the ambient atmosphere.

**Gaseous nucleophilic exchange:** The fluorinated graphene sample (about 1 cm$^2$ on a substrate) was placed in a vial (≈40 mL). The chamber was then evacuated with a membrane pump and refilled with argon (>99.95%) three times to remove oxygen. The nucleophile was then introduced via a Hamilton syringe through a septum on the bottom of the chamber without direct contact with the fluorinated graphene. The reaction thus proceeds only with the nucleophiles in the gas phase. After exposure for 2 h at r.t., the excess of the reagent was removed using a membrane pump and the samples were further exposed to high vacuum ($10^{-5}$ mbar) for 15 min.

**Diazonium grafting:** Grafting was performed according to the published protocols.[6,22,34] Diazonium salts were diluted in deionized water (>18 MΩ cm$^{-1}$) or mixture of deionized water and acetonitrile (7:3) to help the solubility to give 5 mM solution. Graphene on a substrate was immersed into the solution for 2 hours, then removed and thoroughly washed with water and spectroscopy grade methanol.

**Diazonium synthesis:** Commercial diazonia were used if available. Otherwise, they were synthesized according to the following protocol: an aromatic amine (75 μmol) was diluted or suspended in 5 mL of the mixture of deionized water/acetonitrile 7:3 and 1.1 eq. of NaNO$_2$ was added. The reaction mixture was then acidified by 1M HCl to reach pH of ~3-4 when graphene on a substrate was immersed in. Reaction was carried out for 30 min at r.t., then the sample was removed and thoroughly washed with deionized water and methanol (spectroscopy grade).

The successful functionalization of graphene was confirmed by Raman spectroscopy, surface-enhanced Raman spectroscopy, X-ray photoelectron spectroscopy and laser-desorption ionization mass spectrometry and the results are in agreement with the previously reported data.[18]

## 2.2 Infrared spectroscopy

The FTIR spectra of the functionalized graphene samples deposited on the coper foil surface were recorded using ATR (Ge-crystal) and RAIRS techniques on a ThermoScientific Nicolet iN10 FTIR Microscope with a resolution of 4 cm$^{-1}$ and Norton-Beer strong apodization function in the 675–4000 cm$^{-1}$ spectral region, average spot size 150x150 (RAIRS) or 100x100 μm (ATR).

## 2.3 Light and scanning electron microscopy

Surface of copper substrates with functionalized graphene was visualized by light microscopy (LM; microscope DM6000 M; Leica, Austria) and scanning electron microscopy (SEM; microscope Quanta 200 FEG; FEI, Czech Republic). In the light microscope, the samples were observed "as received" using reflected light (episcopic illumination, LM/EPI). In the scanning electron microscope, the samples were fixed with double adhesive carbon tape on a conductive support and observed at accelerating voltage 30 kV using secondary electron imaging (SEM/SE).

## 3 Results and discussion

CVD-grown graphene is typically grown on copper and then transferred onto silicon wafer with silicon dioxide layers (Si/SiO$_2$) which offer superb flatness, good optical contrast and stronger Raman signal than on the original copper foil due to interference and reflection of light from the substrate. However, the transfer procedure introduces contamination to the monolayer from etchant or polymer residuals and brings another variable into the discussion of on-surface covalent chemistry as it has been shown that reactivity of graphene on substrates differs to large extent. Moreover, the Si/SiO$_2$ wafer has its own spectroscopic features which are often overlapping with or even completely jeopardizing all the signal of the grafted species. We have therefore exploited the currently published methodology for covalent functionalization and characterization of CVD graphene[5,18,19,22,35–37] and investigated the potential of infrared spectroscopy to determine the characteristic vibrations of the grafted species. The strict monolayer character of the material is naturally imposing critical challenges on the sensitivity and thus two techniques inherently designed for 2D samples have been used: Attenuated Total Reflection (ATR) and Reflection-Absorption IR Spectroscopy (RAIRS).

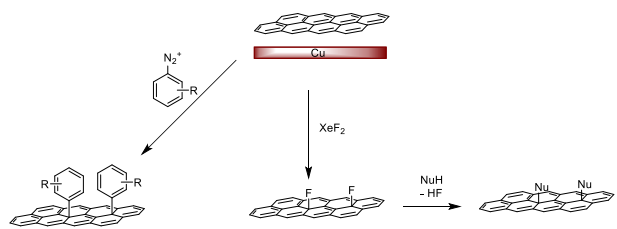

*Figure 1 Schematic reactions used for graphene covalent functionalization.*

Graphene has been grown, functionalized and characterized on copper as described earlier[18] and in the Experimental part. Reaction with diazonia is among the most frequently used functionalization protocols and we have thus taken the 4-nitrophenyldiazonium as benchmark grafting agent for IR investigations, because the strongly electron-withdrawing character of the nitro group prevents cascade azo-coupling reactions leading to polymeric species.[38] Indeed, when graphene on copper was reacted in aqueous diazonium solution, washed and dried, characteristic vibrational bands of nitrophenyl groups have been observed in the IR spectrum (Figure 2), namely 1524 and 1349 cm$^{-1}$ corresponding to asymmetric and symmetric

NO$_2$ stretching vibration, respectively, 1111 cm$^{-1}$ for the C-N stretching, 752 cm$^{-1}$ assigned to NO$_2$ deformation vibrations, and 698 cm$^{-1}$ can be attributed to the NO$_2$ wagging vibration. The signal due to the phenyl groups are indeed detected as well, in particular CH and C=C stretching vibrations at 3084 cm$^{-1}$ and 1598 cm$^{-1}$, respectively. The signal at around 857 cm$^{-1}$ probably overlaps the contribution of CH out-of-plane vibration and NO$_2$ deformation band as it is apparent from its shape. The bands at 2930 and 2860 cm$^{-1}$ were encountered on all graphene samples regardless functionalized or not and probably arise from CH$_2$ stretching vibrations of adsorbed species from ambient atmosphere.

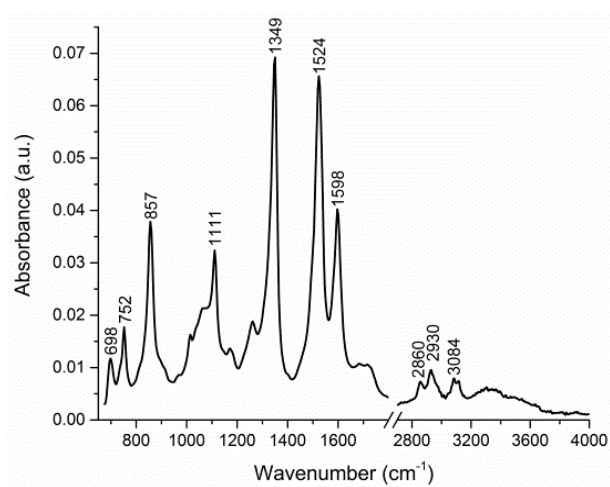

*Figure 2 ATR-IR spectrum of the p-nitrophenyl-functionalized graphene on copper. Bands corresponding to all the characteristic vibrations of the grafted species functional groups are observed: 3084, 1598, 1524, 1349, 1111, 850 and 752 cm$^{-1}$, assignment is provided in the text.*

There are three important aspects concerning the spectrum: a) it is essentially identical those of comparable structural analogues (e.g. *p*-nitrotoluene), b) the same spectrum can be acquired using both RAIRS and ATR technique, and c) all attempts to acquire any spectrum of the same sample on Si/SiO$_2$ substrate provided only very intense bands of SiO$_2$ without any detectable signal of graphene or nitrophenyl moieties.

The results of this experiment encouraged us to explore further the range of detectable functional groups characteristic vibration introduced via covalent grafting. For instance, the Figure 3a shows the spectra of graphene decorated with *p*-ethylcarboxyphenyl moieties measurable by both reflectance and ATR IR techniques. Vibrational manifestations corresponding to the chemical species can be again assigned, in particular the band at 1719 cm$^{-1}$ characteristic for stretching vibration of C=O of an ester carbonyl group, 1279 and 1108 for C-O-C asymmetric and symmetric stretching modes, 1603 cm$^{-1}$ for C=C stretching vibration and also the bands at 1461 and 1365 cm$^{-1}$ that can be assigned to the asymmetric and symmetric deformation vibration of C-H bonds of the ethyl group and band at 1164 and 720 cm$^{-1}$ is associated to the C-C skeletal vibration of alkyl chain. The measured infrared spectrum is again in full agreement with structural analogues. Bands between 2927 and 2855 cm$^{-1}$ corresponding to CH$_3$ and CH$_2$ asymmetric and symmetric stretching vibrations are somewhat hampered by the inherent signals detected in this region on all graphene samples. The band assignment mentioned above is also valid for the *p*-carboxyphenyl functionalization. The FTIR spectrum presented in Figure 3b additionally shows broad absorption bands with maxima in the interval 3200-3400 cm$^{-1}$ which is corresponding to OH stretching vibration from both carboxylic OH groups involved in hydrogen bond system with adsorbed water molecules. The spectra are again very similar by RAIRS and ATR techniques (Figure 3a), the former showing larger contribution of specular reflection components leading to slightly deformed bands which could not have been corrected (due to their mixed origin) even when applying Kramers-Kronig (K-K) transformation. This difference in the character of the spectra leads to the slight shift of the maxima and also influences background shape.

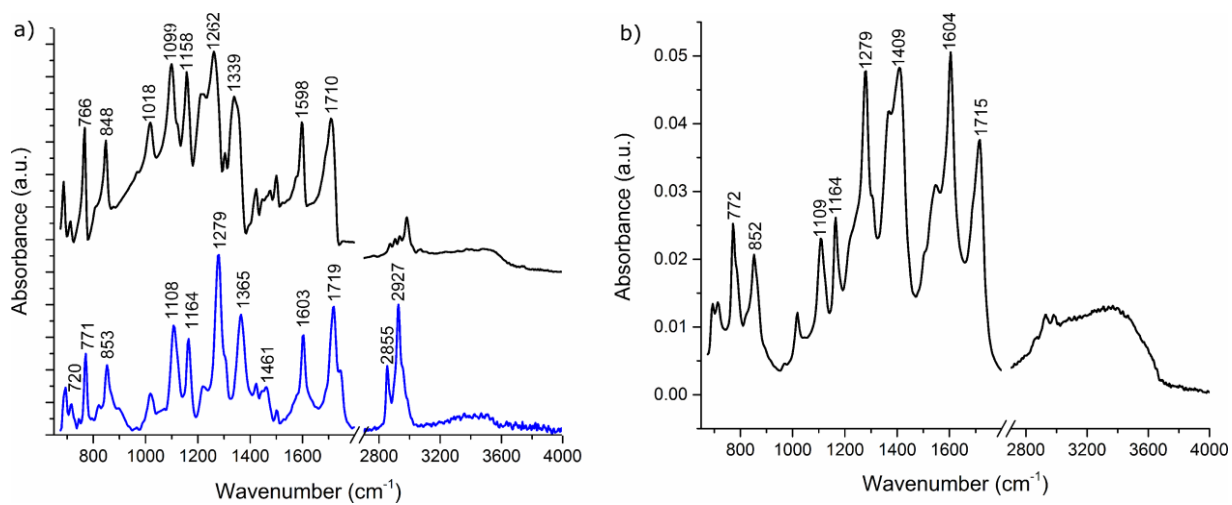

*Figure 3 a) RAIRS (black) and ATR (blue) infrared spectra of 4-ethylcarboxyphenyl functionalized graphene on copper. b) p-Carboxyphenyl functionalized graphene measured in ATR mode. Band assignment for both spectra is given in the text.*

Diazonium salts can in principle undergo two different reaction mechanisms either by radical C-C bond formation (Meerwein arylation) or electrophilic azo-coupling. The latter is particularly effective for diazonia derived from electron-rich aromates ultimately leading to polymerization and even branching.[38] When functionalizing graphene with for instance 4-dimethylaminophenyldiazonium the first radical grafting yields such an electron-rich moiety which then reacts much faster following the azo-coupling mechanism. This is clearly visible in Figure 4 where a series of overlapping bands recorded between 1300 and 1600 cm$^{-1}$ indicate presence of at least two species evidenced for instance by C=C stretching vibration at

1606 and 1590 cm$^{-1}$, broad CH out-of-plane mode at around 850 cm$^{-1}$, broad band at around 1520 cm$^{-1}$ comprising for azo, azoxy and nitroso compounds and dimers (see Figure S1 in the SI) produced by nitrosation of the electron-rich aromate. The bands at 1445 and 1353 cm$^{-1}$ can be assigned to the asymmetric and symmetric bending vibrations of CH$_3$ groups and band at 1164 cm$^{-1}$ corresponds to the rocking CH$_3$ vibration. Additionally, the band at 1353 cm$^{-1}$ corresponds also to the stretching C-N vibration of tertiary aromatic amine. Medium intensity bands at 1230 and 1130 cm$^{-1}$ correspond to mixed asymmetric C-C-N stretching and CH$_3$ rocking modes. Moreover, IR mapping of the surface revealed that the functionalization is largely non-homogeneous, probably due to the fact that after the first radical C-C coupling of the diazonium to graphene (typically slow for electron-rich diazonia if not promoted electrochemically[7]) the following electrophilic azo coupling is much faster leading to polymer growth in the spot and thus jeopardizing the radical mechanism of graphene functionalization (see SI for optical images).

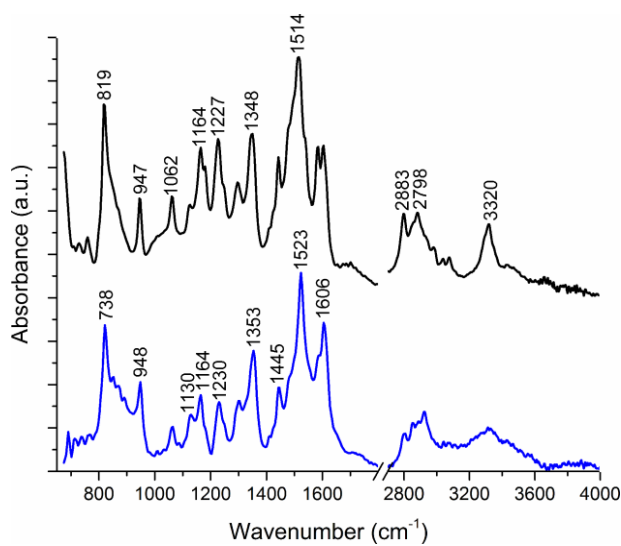

*Figure 4 RAIRS (black) and ATR (blue) spectra of p-(Me$_2$N)-phenyl functionalized graphene are in agreement with the observation of electron-rich diazonia polymerization in functionalization of 2D materials.*

An alternative protocol for covalent graphene functionalization employs primary activation by fluorination[19,39] which is then subjected to nucleophilic substitution by for instance S- or N- nucleophiles replacing fluorine with such moieties. Unlike in the diazonium reaction, formation of assembled layers on bare substrate and/or physical adsorbates on graphene cannot be *a priori* excluded, and therefore in the case of nucleophilic exchange, it is necessary to compare three IR spectra: bare copper substrate exposed to the nucleophile, graphene on copper without fluorination exposed to the nucleophiles and finally, the functionalized material by substitution of fluorinated graphene (Figure 5). When thiophenol has been used the nucleophilic agent, no signal corresponding to a characteristic vibration could have been observed in the spectra nor on bare copper nor graphene. However, when fluorinated graphene on copper has been reacted with thiophenol, series of sharp bands appeared: aromatic C-H stretch at 3065 cm$^{-1}$, C=C stretching mode at 1579 cm$^{-1}$, C-H deformation vibration at 1476 and 1439 cm$^{-1}$, combined aromatic CH in-plane vibrations with C-C ring vibration at 1081, 1024 and 1000 cm$^{-1}$, as mixed CH out-of-plane and C-S stretching vibrations at 738 and 690 cm$^{-1}$ characteristic for monosubstituted benzene ring. Importantly, the acquired spectra are basically identical to those of thiophenol with the exception of the bands of S-H stretching and deformation modes at 2550 and 900 cm$^{-1}$, respectively, which are not present in the spectrum thus clearly confirming the nucleophilic exchange reactional pathway. Very similar results were obtained also for benzylthiol (see SI). The spectra of reactions with thioacetic acid show that the bare substrate is indeed corroded by the action of the reagent (like in the case of propylamine, see SI) as evidenced by the characteristic CO-S vibration at 872 cm$^{-1}$. Moreover, it is observed that significant amount of sulfur atoms are oxidized to sulfoxide (1099 cm$^{-1}$) or even sulfone state (1419 and 1349 cm$^{-1}$ for asymmetric and symmetric SO$_2$ vibration respectively) which has been also observed in XPS and TPD data earlier.[18] It can be also observed than the tautomeric equilibrium between (C=O)-S and (C=S)-O states leads to broadening of the carbonyl band at around 1650 cm$^{-1}$ and appearance of the C=S vibration at 1136 cm$^{-1}$.

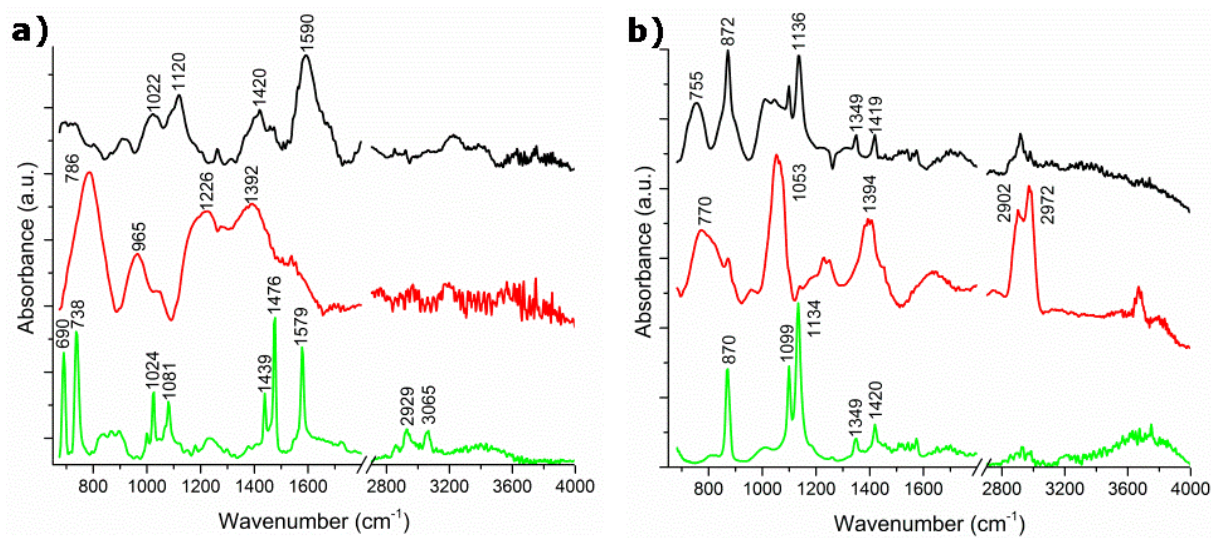

*Figure 5 a) Partly fluorinated graphene on copper reacted with thiophenol (green trace), and comparison with two "blank" experiments – bare substrate exposed to the nucleophile (black) and graphene on copper exposed to thiophenol (red). The characteristic signals corresponding to the phenylsulfanyl moiety are only observed on fluorinated graphene. b) Reaction of fluorographene with thioacetic acid on copper (green) and comparison with blank experiments on bare substrate (black) and non-fluorinated graphene (red). Thioacetic acid is heavily corrosive for the copper foil as evidenced by appearance of*

*characteristic vibrational modes on bare substrate which are largely suppressed when graphene is present on its surface. Nucleophilic exchange on fluorographene gives intense bands characteristic for the thioacetyl moiety (assignment given in the text).*

It has to be noted that in none of the above mentioned cases any signal was detected when measuring on Si/SiO$_2$ substrate on which the spectra were totally dominated by the strong absorption of SiO$_2$ layer. This is also in accord with the existence of the surface plasmons of the copper substrate which has been previously identified responsible for enhancing infrared spectra.[25–29] The enhancement is the highest for vibrations with perpendicular dipole moment change with respect to the metal surface, which changes the relative band intensities in the spectra as compared with the parent compounds. In the case of nucleophilic substitutions, the possibility of copper corrosion has to be considered, but as shown by comparison of spectra with bare substrate, graphene on substrate and fluorinated graphene on copper it is not responsible for the appearance of characteristic vibrations signal on the covalently grafted material. Both LM and SEM showed the characteristic surface morphology due to copper sheet rolling and some locations on the sample also displayed Cu microcrystal grains, as summarized in Figure S4 in the SI. Importantly, the surface corrosion is producing only microscopic asperities which are most likely beneficial for the enhancement of spectroscopic signal.

## 4 Conclusions

We have demonstrated the use of infrared spectroscopy in the characterization of series of covalent functionalization of CVD-grown monolayer graphene either by radical mechanism employing diazonium salts (Meerwein arylation) or by exchange of partly fluorinated graphene for thiols or amines (nucleophilic substitution). Characteristic vibrations of functional groups attached to graphene could have been detected and assigned. The spectra were measured with reflectance by RAIRS and ATR techniques without significant changes in the obtained data. The signal enhancement is due to surface plasmons of the copper substrate which is used for synthesis of CVD graphene and hence no signal could have been recorded on e.g. Si/SiO$_2$ wafers. Infrared spectroscopy has a great potential for being applied in characterization of graphene functionalization due to the low-energy radiation and non-destructive operation, making it valuable alternative to Raman spectroscopy routinely used in this field.

## Acknowledgements

The work was supported by Czech Science Foundation contract No. 15-01953S and ERC-CZ project No. LL1301. P.K. thanks the ASCR PPPLZ program for funding (L200401551). The authors acknowledge the assistance provided by the Research Infrastructures NanoEnviCz (Project No. LM2015073) and Pro-NanoEnviCz (Project No. CZ.02.1.01/0.0/0.0/16_013/0001821), supported by the Ministry of Education, Youth and Sports of the Czech Republic. Authors thank Karolina A. Drogowska for graphene synthesis.

## Supplementary information

Spectra of additional functionalizations, optical images and supporting figures can be found in the Supplementary information file.